\documentclass[seceq,twocolumn]{jpsj3}
\usepackage{txfonts}
\usepackage{bm}
\usepackage{braket}
\usepackage{color}
\newcommand{\ds}{\displaystyle}

\title{Competing Kondo Effects in Non-Kramers Doublet Systems}

\author{Hiroaki \textsc{Kusunose}\thanks{E-mail address: hk@meiji.ac.jp}
}

\inst{Department of Physics, Meiji University, Kanagawa 214-8571, Japan
}

\abst{
In non-Kramers Kondo systems with a quadrupolar degrees of freedom, an ordinary magnetic Kondo effect can compete with the quadrupolar Kondo effect.
We discuss such competition keeping Pr$T_{2}$Zn$_{20}$ ($T$=Ir, Rh) and Pr$T_{2}$Al$_{20}$ ($T$=V, Ti) in mind, where the $\Gamma_{3}$ non-Kramers crystalline-electric-field (CEF) doublet ground state is realized in Pr$^{3+}$ ion with $(4f)^{2}$ configuration under cubic symmetry.
The quadrupolar Kondo effect can be described by the two-channel Kondo model, which leads to the local non-Fermi-liquid (NFL) ground state, while the magnetic Kondo effect favors the ordinary local Fermi-liquid (FL) ground state.
Based on the minimal extended two-channel Kondo model including the magnetic Kondo coupling as well, we investigate the competition and resulting thermodynamics, and orbital/magnetic, and single-particle excitation spectra by the Wilson's numerical renormalization-group (NRG) method.
There exists the first-order transition between the NFL and FL ground states.
In addition to these two states, the alternative FL state accompanied by a free magnetic spin appears in intermediate temperature range, which eventually reaches the true NFL ground state, as a consequence of stronger competition between the magnetic and quadrupolar Kondo effects.
In this peculiar state, the magnetic susceptibility shows a Curie-like behavior, while the orbital fluctuation exhibits the FL behavior. Moreover, the single-particle spectra yield a more singular behavior. Implications to the Pr 1-2-20 systems are briefly discussed.
}

\begin{document}
\maketitle

\section{Introduction}

Fundamental aspect of Ce-based heavy-fermion systems has been understood by investigating magnetic Kondo effect, by which the large entropy of the localized $f$ electrons at high temperatures is released to form Landau quasiparticles with the heavy effective mass.
The Kondo effect also gives rise to the indirect magnetic interaction via conduction electrons, which suppresses the Kondo screening and leads to magnetically ordered state.
The competition between screening and stabilizing the $f$-electron magnetic moments leads to the quantum critical point (QCP), and a variety of superconductivities has emerged in the vicinity of the QCP.

In contrast to the above ``conventional'' view of heavy fermions that is well known as the Doniach picture in the Ce-based systems, Pr-based and U-based compounds require a different picture as they are characterized by plural $f$-electron configuration, which allows us to explore pure orbital physics with non-Kramers degeneracy.
In such context, recently found Pr$T_{2}$Zn$_{20}$ ($T$=Ir, Rh) and Pr$T_{2}$Al$_{20}$ ($T$=V, Ti) provide interesting playgrounds since they are characterized by the non-Kramers $\Gamma_{3}$ doublet CEF ground state, in which electric quadrupoles and a magnetic octupole become active but no magnetic dipoles play a role in low-energy physics.~\cite{Onimaru10,Onimaru11,Sakai11,Ishii11,Sakai12,Sato12,Onimaru12,Iwasa13,Tsujimoto14,Araki14,Kuramoto09}
Indeed, they show considerable deviation in temperature dependences of the specific heat, resistivity and so on from the conventional FL behaviors.~\cite{Sakai11,Ishii11,Ikeura14,Machida15}
They also exhibit the electric quadrupole order which even coexists with superconductivity.~\cite{Onimaru10,Onimaru11,Sakai11,Ishii11,Koseki11,Sakai12,Onimaru12,Ishii13,Tsujimoto14,Araki14,Matsumoto15,Matsubayashi12,Matsubayashi14,Hattori14}

The Al-systems show a logarithmic increase in the resistivity at high temperatures,~\cite{Sakai11} which is an indication of the magnetic Kondo effect, while the Zn-systems exhibit no magnetic Kondo effects.~\cite{Onimaru10,Onimaru11}
Moreover, the conventional $T^{2}$ behavior is observed in the resistivity of PrTi$_{2}$Al$_{20}$ at ambient pressure,~\cite{Sakai11} which drastically changes to anomalous behaviors by applying pressure.~\cite{Sakai11,Matsubayashi12,Matsubayashi14}

Motivated by these observations, an interplay between the conventional magnetic Kondo effect and electric quadrupolar Kondo effect in the non-Kramers doublet systems is addressed.~\cite{Kusunose15}
As a consequence of the competition between two Kondo effects, the peculiar electronic state emerges in which the magnetic susceptibility shows a Curie-like behavior, while the quadrupolar response is the FL-like.
The resistivity has a more anomalous $T$ dependence than the NFL~\cite{Cox98,Emery92,Affleck95,Tsvelick85,Sacramento89,Sakai93,Cox93,Jarrell96,Cox93a,Cox87,Hoshino11,Hoshino13,Hoshino14,Hoshino14a,Tsuruta99,Tsuruta00,Tsuruta15} of the two-channel Kondo model, $T^{1/2}$.
This peculiar state is very sensitive to a detailed balance of the two Kondo effects, which implies a relevance to the enigmatic behaviors of PrTi$_{2}$Al$_{20}$ under pressures.~\cite{Sakai11,Matsubayashi12,Matsubayashi14}

This paper is organized as follows.
In the next section, we introduce the minimal extended two-channel Kondo model in terms of the simplified $j$-$j$ coupling scheme.
We identify two limiting cases which correspond to the FL and NFL fixed points.
In \S3, using the NRG calculation,~\cite{Wilson75,Bulla08} we present the overall phase diagram, thermodynamic properties and dynamical spectral functions.
Based on these results, we discuss the properties of the peculiar electronic state that is emerged by the strong competition between two Kondo effects.
The final section summarizes the paper.

\section{Extended Two-Channel Kondo Model}

\subsection{Minimal model}

The magnetic ion, Pr$^{3+}$, in Pr$T_{2}$Zn$_{20}$ ($T$=Ir, Rh) and Pr$T_{2}$Al$_{20}$ ($T$=V, Ti) has $(4f)^{2}$ configuration with the total angular momentum, $J=4$.
Under the cubic $T_{d}$ ($T$ in the case of PrRh$_{2}$Zn$_{20}$) point-group symmetry, $J=4$ multiplet splits into $\Gamma_{1}$ singlet, $\Gamma_{3}$ doublet, and two $\Gamma_{4}$, $\Gamma_{5}$ triplets.
The non-Kramers degeneracy appears due to even number of localized $f$ electrons, to which the Kramers theorem cannot apply.
The CEF states in these compounds are commonly characterized by the $\Gamma_{3}$ non-Kramers doublet ground state and the first-excited $\Gamma_{4}$ or $\Gamma_{5}$ magnetic triplet state.~\cite{Onimaru10,Onimaru11,Sakai11,Ishii11,Sakai12,Sato12,Onimaru12,Iwasa13,Tsujimoto14,Araki14}

The non-Kramers $\Gamma_{3}$ doublet has the pure electric quadrupoles, and a magnetic octupole.
Since the localized degrees of freedom is non-magnetic, there exist two equivalent scattering channels corresponding to the time-reversal pair of conduction electrons with quadrupoles.
This is a heart of the so-called quadrupolar Kondo effect that is described by the two-channel Kondo model.
Its ground state is known as the local NFL due to overscreening of the $f$-electron quadrupole by two quadrupoles of the conduction electrons with the $\uparrow$ and $\downarrow$ spins.~\cite{Cox98}
On the other hand, the excited triplet state has magnetic dipole moments, which are screened by an ordinary magnetic Kondo effect, leading to the local FL ground state.

In order to describe these two Kondo screening processes, we construct a minimal Kondo impurity system with the electric quadrupoles and magnetic dipoles by adopting the $j$-$j$ coupling scheme,~\cite{Kusunose05} where $f^{2}$ states can be expressed by putting two electrons into one-body $f$-electron states, $\Gamma_{7}$ doublet ($f_{7\sigma}^{\dagger}$ with $\sigma=\uparrow,\downarrow$), and $\Gamma_{8}$ quartet ($f_{m\sigma}^{\dagger}$ with $m=1,2$).
In what follows, we have neglected the doubly occupied states, $\ket{m\uparrow}\ket{m\downarrow}$ ($m=1,2,7$), by assuming the large intra-atomic Coulomb repulsion for simplicity.

Introducing the spin operators,
\begin{equation}
\bm{S}_{fm}=\sum_{\alpha\beta}f_{m\alpha}^{\dagger}\left(\frac{\bm{\sigma}_{\alpha\beta}}{2}\right)f_{m\beta}^{},
\end{equation}
the $\Gamma_{4}$($\Gamma_{5}$) state can be expressed by the triplet state of the total spin, $\bm{S}_{f1(2)}+\bm{S}_{f7}$, while the singlet states of it constitute the non-Kramers $\Gamma_{3}$ doublet.
The orbital operator can be expressed by
\begin{equation}
\bm{T}_{f}=\sum_{\sigma}\sum_{mn}^{1,2}f_{m\sigma}^{\dagger}\left(\frac{\bm{\tau}_{mn}}{2}\right)f_{n\sigma}^{}.
\end{equation}
Here, $\bm{\sigma}$ and $\bm{\tau}$ are the Pauli matrices acting onto the Kramers and non-Kramers pairs, respectively.
Note that $\bm{T}_{f}$ can flip the ``orbital'' index $m=1,2$, which exchanges the states among the non-Kramers doublet or the $\Gamma_{4}$ and $\Gamma_{5}$ states without changing their spin states.

\begin{figure}[t]
\centering{
\includegraphics[width=3.8cm]{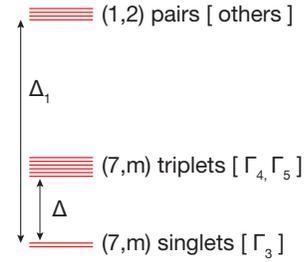}
}
\caption{(Color online) CEF levels of $(4f)^{2}$ configuration in the $j$-$j$ coupling scheme. The doubly occupied states have been neglected due to large intra-atomic Coulomb repulsion.}
\label{f2cef_level}
\end{figure}

In terms of these operators, the minimal extended two-channel Kondo model is given by
\begin{equation}
H=H_{c}+H_{f}+J\sum_{m}^{1,2}\bm{s}_{m}\cdot\bm{S}_{fm}+K\sum_{\sigma}\bm{t}_{\sigma}\cdot\bm{T}_{f},
\end{equation}
where $\bm{s}_{m}$ and $\bm{t}_{\sigma}$ are the spin and orbital operators of the conduction electrons at the impurity site, which are given by
\begin{equation}
\bm{s}_{m}=\sum_{kk'}\sum_{\alpha\beta}c_{km\alpha}^{\dagger}\left(\frac{\bm{\sigma}_{\alpha\beta}}{2}\right)c_{k'm\beta}^{},
\end{equation}
and
\begin{equation}
\bm{t}_{\sigma}=\sum_{kk'}\sum_{mn}^{1,2}c_{km\sigma}^{\dagger}\left(\frac{\bm{\tau}_{mn}}{2}\right)c_{k'n\sigma}^{}.
\end{equation}
The spin exchange comes from a process, e.g.,
\begin{equation}
(f_{1\uparrow}^{\dagger},f_{7\downarrow}^{\dagger};c_{1\downarrow}^{\dagger})
\,\,\,\Rightarrow\,\,\,
(f_{7\downarrow}^{\dagger};c_{1\downarrow}^{\dagger},c_{1\uparrow}^{\dagger})
\,\,\,\Rightarrow\,\,\,
-(f_{1\downarrow}^{\dagger},f_{7\downarrow}^{\dagger};c_{1\uparrow}^{\dagger}),
\end{equation}
while the orbital exchange comes from a process, e.g.,
\begin{equation}
(f_{1\uparrow}^{\dagger},f_{7\downarrow}^{\dagger};c_{2\uparrow}^{\dagger})
\,\,\,\Rightarrow\,\,\,
(f_{7\downarrow}^{\dagger};c_{2\uparrow}^{\dagger},c_{1\uparrow}^{\dagger})
\,\,\,\Rightarrow\,\,\,
-(f_{2\uparrow}^{\dagger},f_{7\downarrow}^{\dagger};c_{1\uparrow}^{\dagger}),
\end{equation}
via the hybridization term, $\varv_{8}\sum_{k}\sum_{m}^{1,2}(f_{m\sigma}^{\dagger}c_{km\sigma}^{}+{\rm h.c.})$.
Note that these processes have the same order of magnitude, i.e., $J$, $K\sim \varv_{8}^{2}/\Delta E$, where $\Delta E$ is the energy difference to the intermediate state.
The kinetic term of the conduction electrons is given by
\begin{equation}
H_{c}=\sum_{k\sigma}\sum_{m}^{1,2}\epsilon_{k}c_{km\sigma}^{\dagger}c_{km\sigma}^{},
\end{equation}
and the CEF splitting is described by
\begin{equation}
H_{f}=\Delta\sum_{m}^{1,2}P_{m}+\Delta_{1}\,n_{f1}n_{f2}.
\end{equation}
Here, $P_{m}=(3/4)n_{fm}n_{f7}+\bm{S}_{fm}\cdot\bm{S}_{f7}$ with the $f$-electron number in the $m$ orbital, $n_{fm}=\sum_{\sigma}f_{m\sigma}^{\dagger}f_{m\sigma}^{}$, is the projection operator onto the two triplet states.
We have also introduced the other excited state $\Delta_{1}$, in which two $f$ electrons occupy the $m=1$ and $m=2$ orbitals with leaving the $\Gamma_{7}$ state empty.
We have neglected the energy splittings between the $\Gamma_{4}$ and $\Gamma_{5}$ states, and among the pair-electron states in the $\Gamma_{8}$ orbitals for simplicity.
The CEF levels in the $j$-$j$ coupling scheme and their expressions are shown in Fig.~\ref{f2cef_level} and Table~\ref{tbl1}.

\begin{table}[t]
\caption{\label{tbl1}$f^{2}$ CEF states in the $j$-$j$ coupling scheme. $S_{f}$ is the magnitude of the total $f$ spin, $\bm{S}_{f}=\sum_{m}^{1,2,7}\bm{S}_{fm}$. There are $_{6}C_{2}=15$ states in total.}
\begin{center}
\begin{tabular}{cccclc}
\hline
CEF & $S_{f}$ & deg. & energy & eigenstate \\
\hline
``$\Gamma_{3}$'' & $0$ & $1$ & $0$ & $\ds\frac{1}{\sqrt{2}}(\ket{1,\uparrow}\ket{7,\downarrow}-\ket{1,\downarrow}\ket{7,\uparrow})$ \\
             &     & $1$ &     & $\ds\frac{1}{\sqrt{2}}(\ket{2,\uparrow}\ket{7,\downarrow}-\ket{2,\downarrow}\ket{7,\uparrow})$ \\
\hline
``$\Gamma_{4}$'' & $1$ & $3$ & $\Delta$  & $\ket{1,\downarrow}\ket{7,\downarrow}$ \\
             &     &     &           & $\ds\frac{1}{\sqrt{2}}(\ket{1,\uparrow}\ket{7,\downarrow}+\ket{1,\downarrow}\ket{7,\uparrow})$ \\
             &     &     &           & $\ket{1,\uparrow}\ket{7,\uparrow}$ \\
``$\Gamma_{5}$'' &     & $3$ &           & $\ket{2,\downarrow}\ket{7,\downarrow}$ \\
             &     &     &           & $\ds\frac{1}{\sqrt{2}}(\ket{2,\uparrow}\ket{7,\downarrow}+\ket{2,\downarrow}\ket{7,\uparrow})$ \\
             &     &     &           & $\ket{2,\uparrow}\ket{7,\uparrow}$ \\
\hline
pair in $\Gamma_{8}$ & $1$ & 3  & $\Delta_{1}$ & $\ket{1,\downarrow}\ket{2,\downarrow}$ \\
       &     &    &              & $\ds\frac{1}{\sqrt{2}}(\ket{1,\uparrow}\ket{2,\downarrow}+\ket{1,\downarrow}\ket{2,\uparrow})$ \\
       &     &    &              & $\ket{1,\uparrow}\ket{2,\uparrow}$ \\
       & $0$ & 1  &              & $\ds\frac{1}{\sqrt{2}}(\ket{1,\uparrow}\ket{2,\downarrow}-\ket{1,\downarrow}\ket{2,\uparrow})$ \\
\hline
doubly   & $0$ & 1  & $\infty$ & $\ket{1,\downarrow}\ket{1,\uparrow}$ \\
occupied &     & 1  &          & $\ket{2,\downarrow}\ket{2,\uparrow}$ \\
         &     & 1  &          & $\ket{7,\downarrow}\ket{7,\uparrow}$ \\
\hline
\end{tabular}
\end{center}
\end{table}

\subsection{Quadrupolar Kondo limit}

Let us consider the case where the excited CEF states can be neglected, or $K\gg J$.
In this case, $\bm{S}_{fm}$ becomes inactive, and within the $\Gamma_{3}$ doublet, $\ket{u}$ and $\ket{\varv}$, the orbital operator is reduced to
\begin{equation}
\bm{T}_{f}\to \bm{\tau}_{f}\equiv\sum_{\mu\nu}^{u,\varv}f_{\mu}^{\dagger}\left(\frac{\bm{\tau}_{\mu\nu}}{2}\right)f_{\nu}^{}.
\end{equation}
Then, the extended two-channel Kondo model is reduced to the ordinary two-channel Kondo model as
\begin{equation}
H=H_{c}+K\sum_{\sigma}\bm{t}_{\sigma}\cdot\bm{\tau}_{f},
\end{equation}
and we expect the NFL ground state in this limit.~\cite{Cox98,Emery92,Affleck95,Tsvelick85,Sacramento89,Sakai93,Cox93,Jarrell96,Cox93a,Cox87}

\subsection{Magnetic Kondo limit}

The other limit is $J\gg K$, $\Delta_{1}$.
In this case, both $f$ electrons tend to occupy $\Gamma_{8}$ states in order to gain the magnetic exchange energy, $J$, which correspond to the excited states with the energy $\Delta_{1}$.
Since $\bm{S}_{f7}$ and $\bm{T}_{f}$ become inactive in this case, the model is reduced to
\begin{equation}
H=H_{c}+J\sum_{m}^{1,2}\bm{s}_{m}\cdot\bm{S}_{fm}.
\end{equation}
This is nothing but two independent single-channel Kondo models, yielding the ordinary FL behaviors.

\section{Results}

We perform the NRG calculation on the extended two-channel Kondo model.
In the NRG computation,~\cite{Bulla08} the conduction band is discretized logarithmically to focus on the low-energy excitations.
For this purpose, the kinetic-energy term is transformed to the hopping-type Hamiltonian on the semi-infinite one-dimensional chain, which can be solved by iterative diagonalization with appropriate truncation of higher-energy states.
In this paper, we have used the discretization parameter $\Lambda=3$, and retained at least 700 low-energy states for each iteration step, $N$.
Note that the iteration step $N$ corresponds to the energy (temperature) scale of $\Lambda^{-(N-1)/2}$, and hence the larger $N$ corresponds to lower $T$.
The CEF splittings are chosen as $\Delta=1\times10^{-3}$ and $\Delta_{1}=5\times10^{-3}$ in unit of the half bandwidth of the conduction electrons, $D=1$, throughout this paper.

Since we neglect the hybridization for the $\Gamma_{7}$ orbital, the sectors with $n_{f7}=0$ and $n_{f7}=1$ are decoupled, and $n_{f7}$ does not change for whole iteration steps.
Note that the quadrupolar (magnetic) Kondo limit belongs to $n_{f7}=1$ ($n_{f7}=0$) sector.
We can keep track of the low-energy states by two conversed quantities in addition to $n_{f7}$, namely, $\mathcal{C}\equiv(Q,S_{z},n_{f7})$, where $Q$ is the total electron number measured from the half-filling and $S_{z}$ is the $z$-component of the total spin.
The total number of $f$ electrons is kept fixed as $n_{f}=2$.

\subsection{Phase diagram}

\begin{figure}[t]
\centering{
\includegraphics[width=8.5cm]{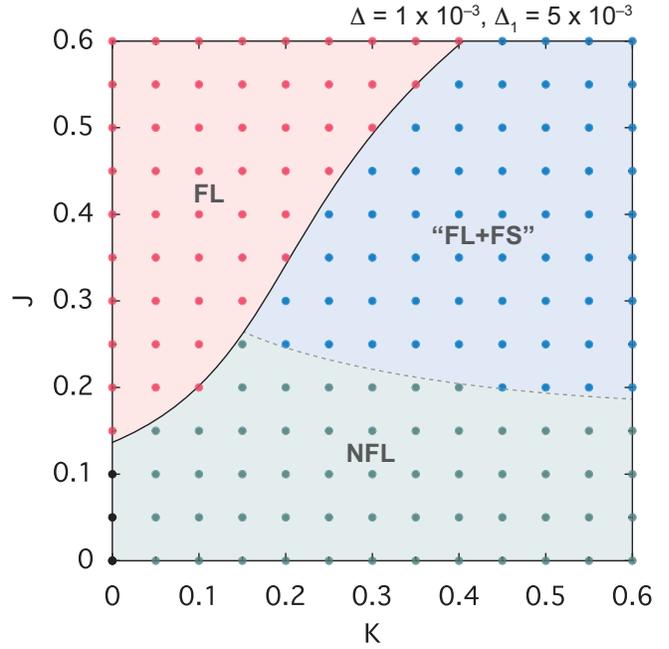}
}
\caption{(Color online) Phase diagram for the extended two-channel Kondo model. FL(NFL) indicates the local Fermi (non-Fermi) liquid ground state. ``FL+FS'' represents an alternative Fermi-liquid state with effective free magnetic spin $S=1/2$ at the lowest temperature we computed. The true ground state eventually becomes NFL. The black circles at $K=0$ indicate the CEF doublet ground state decoupled with the conduction electrons.}
\label{phase}
\end{figure}

Figure~\ref{phase} shows the phase diagram of the extended two-channel Kondo model.
There are three distinct states at the lowest $T$, two of which are nothing but the states we expected in the limiting cases.
Namely, the larger $J$ region is the FL ground state that is characterized by $\mathcal{C}=(0,0,0)$, while the larger $K$ region is the NFL ground state with doubly degenerate, $\mathcal{C}=(0,0,1)^{2}$.
For $K=0$ and the smaller $J$, we have the isolated CEF doublet ground state decoupled with the conduction electrons.
The boundary between the NFL and FL at $K=0$ is roughly determined by the condition that the magnetic Kondo temperature, $T_{\rm Km}$, is equal to $\Delta_{1}/2$ (See also in Fig.~\ref{TK}).
When the former exceeds the latter, the FL state is stabilized by exciting two $f$ electrons into the $\Gamma_{8}$ states to form two independent Kondo singlets.
Since the FL and NFL ground states have different $n_{f7}$, the transition between them becomes first order.

\begin{figure}[t]
\centering{
\includegraphics[width=8.5cm]{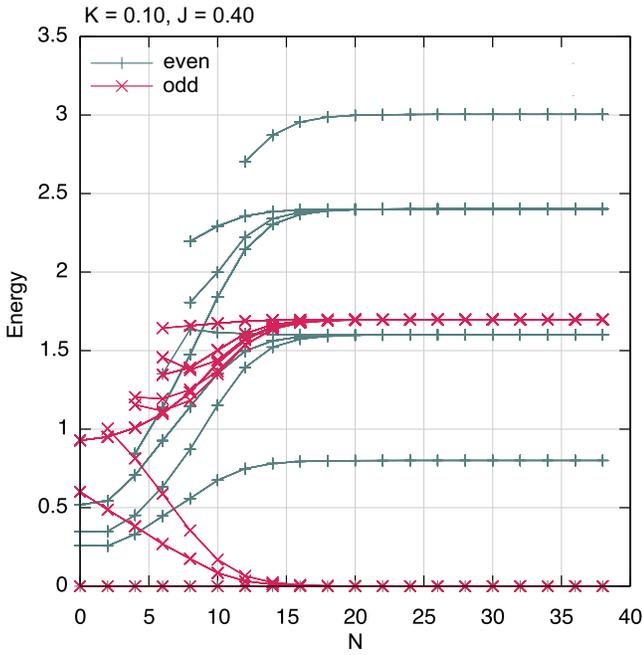}
}
\caption{(Color online) Flow diagram for $K=0.10$, $J=0.40$. The ground state for even $N$ is characterized by $\mathcal{C}=(0,0,0)$, while for odd $N$ the ground states are given by $\mathcal{C}=(0,0,0)^{4}$, $(0,\pm1,0)$, $(\pm2,0,0)$, $(\pm1,\pm1/2,0)^{2}$.}
\label{k010j040}
\end{figure}

\begin{figure}[t]
\centering{
\includegraphics[width=8.5cm]{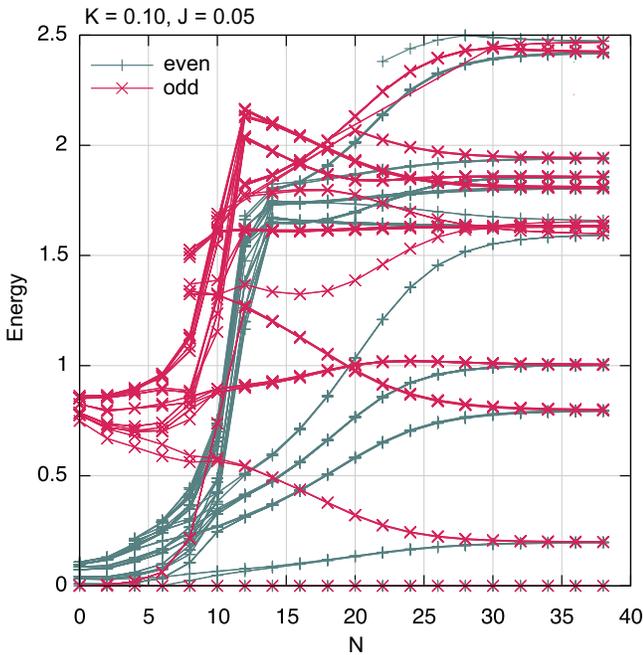}
}
\caption{(Color online) Flow diagram for $K=0.10$, $J=0.05$. The ground states for even and odd $N$ is characterized by doubly degenerate $\mathcal{C}=(0,0,1)^{2}$.}
\label{k010j005}
\end{figure}

The typical flow diagram of the low-lying energies for $K=0.10$, $J=0.40$ at the iteration step, $N$, is shown in Fig.~\ref{k010j040}, where the energies are enlarged by a factor $\Lambda^{(N-1)/2}$.
This spectrum can be interpreted as the individual particle-hole excitations of the ``quasi particles'' on a finite chain, and hence it represents the FL fixed point.~\cite{Wilson75}

On the contrary, the flow diagram for $K=0.10$, $J=0.05$ is shown in Fig.~\ref{k010j005}, which represents the typical excitation spectra of the overscreening NFL fixed point.~\cite{Cox98}
Namely, there is no even-odd alternation, and the spectra cannot be interpreted as simple particle-hole excitations of ``quasi particles''.
Indeed, this excitation spectra including degeneracy coincide with those of the two-channel Kondo model.
Since the $f$-electron configuration differs in the FL and NFL phases, the strong orbital fluctuation between $\Gamma_{7}$ and $\Gamma_{8}$ may be expected near the FL-NFL boundary.
This is the important consequence of the cubic symmetry and the non-Kramers CEF ground state.

\begin{figure}[t]
\centering{
\includegraphics[width=8.5cm]{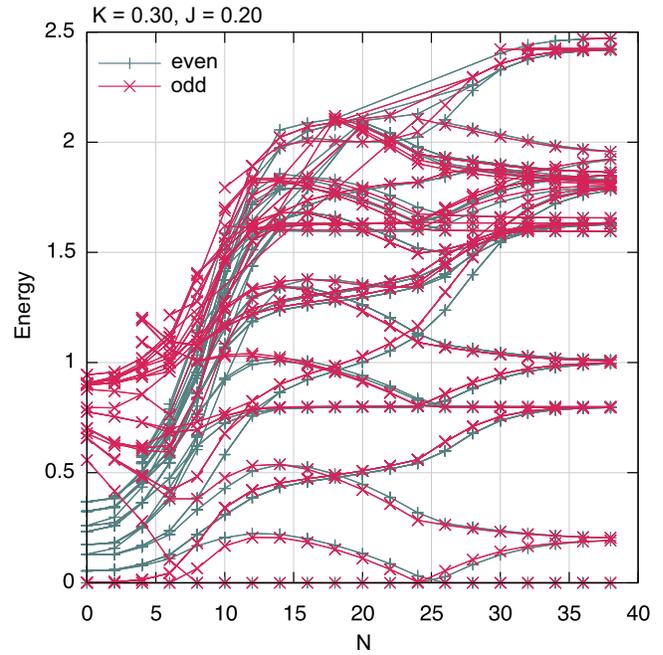}
}
\caption{(Color online) Flow diagram for $K=0.30$, $J=0.20$. The true fixed point is characterized by the NFL excitations as shown in Fig.~\ref{k010j005}. However, an alternative excitation spectra appear at the intermediate energy scales, $N\lesssim 25$, with the ground state $\mathcal{C}=(1,\pm1/2,1)$ for even $N$ and $(-1,\pm1/2,1)$ for odd $N$.}
\label{k030j020}
\end{figure}

The interesting behavior appears for larger $K$ and $J$.
In the flow diagram for $K=0.30$, $J=0.20$ shown in Fig.~\ref{k030j020}, the excitation spectra at the lowest energy scale, i.e., $N=38$ and $39$, are equivalent to those of Fig.~\ref{k010j005}, indicating that the ground state is characterized by the NFL fixed point.
However, at the intermediate energy scales, $N\lesssim N^{*}$ ($N^{*}=25$ for the present parameter set), an alternative excitation spectra appear with the ground state $\mathcal{C}=(1,\pm1/2,1)$ for even $N$ and $(-1,\pm1/2,1)$ for odd $N$.
The crossover step $N^{*}$ becomes larger (the crossover energy scale becomes smaller) as $K$ and $J$ become larger due to the stronger competition between the quadrupolar and magnetic Kondo effects.
As will be shown later, the competing region (indicated by ``FL+FS'' in Fig.~\ref{phase}) is characterized by an effective $S=1/2$ free spin which couples marginally with conduction electrons, as similar to the ferromagnetic or underscreening case of the Kondo effect.

\begin{figure}[t]
\centering{
\includegraphics[width=8.5cm]{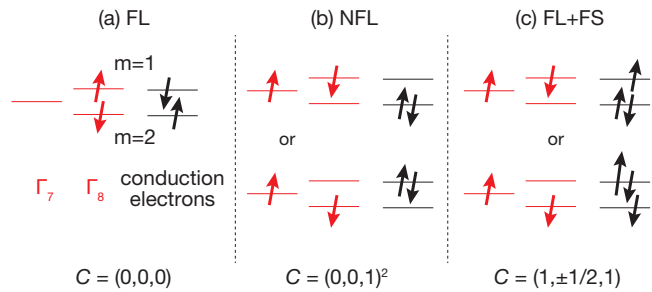}
}
\caption{(Color online) Schematic illustrations of the fixed points, (a) perfect screening FL, (b) the overscreening NFL, and (c) partial screening FL+FS. Note that (a) and (b) are non-magnetic, while (c) is magnetic.}
\label{fp}
\end{figure}

The schematic illustrations of these three fixed points are given in Fig.~\ref{fp}.

\subsection{Thermodynamic properties}

\begin{figure}[t]
\centering{
\includegraphics[width=8.5cm]{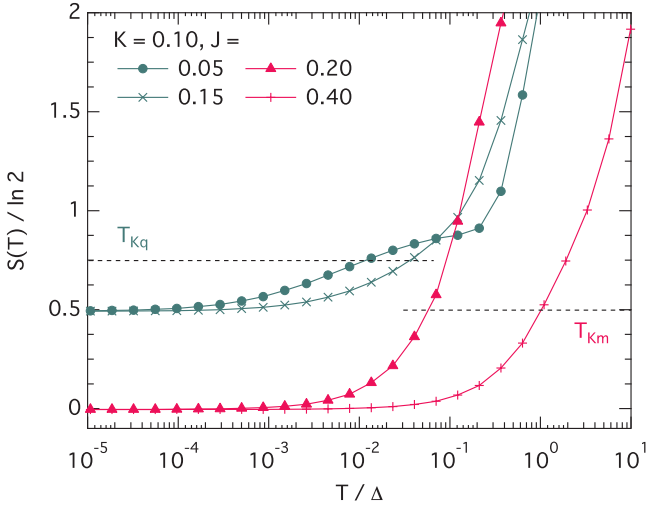}
}
\caption{(Color online) $T$ dependence of the impurity entropy for $K=0.10$. In the NFL phase, we have the residual entropy $(1/2)\ln2$, while in the FL phase, the entropy vanishes. We define the Kondo temperatures for the quadrupolar and magnetic Kondo effects by $S(T_{\rm Kq})=(3/4)\ln2$ and $S(T_{\rm Km})=(1/2)\ln2$, respectively.}
\label{S_K010}
\end{figure}

Next, we discuss the thermodynamic properties of the model.
Thermodynamic quantities are evaluated at the temperature, $T=t\,\Lambda^{-(N-1)/2}$ with $t=0.4$ for each step $N$.
Figure~\ref{S_K010} shows the $T$ dependence of the impurity entropy for $K=0.10$ with increasing $J=0.05$, $0.15$, $0.20$ and $0.40$.
In the NFL phase, we have the residual entropy $(1/2)\ln2$ which is the characteristic feature of the overscreening fixed point of the two-channel Kondo model.
On the other hand, the entropy vanishes in the FL phase.
We define the Kondo temperatures for the quadrupolar and magnetic Kondo effects by $S(T_{\rm Kq})=(3/4)\ln2$ and $S(T_{\rm Km})=(1/2)\ln2$, respectively.

\begin{figure}[t]
\centering{
\includegraphics[width=8.5cm]{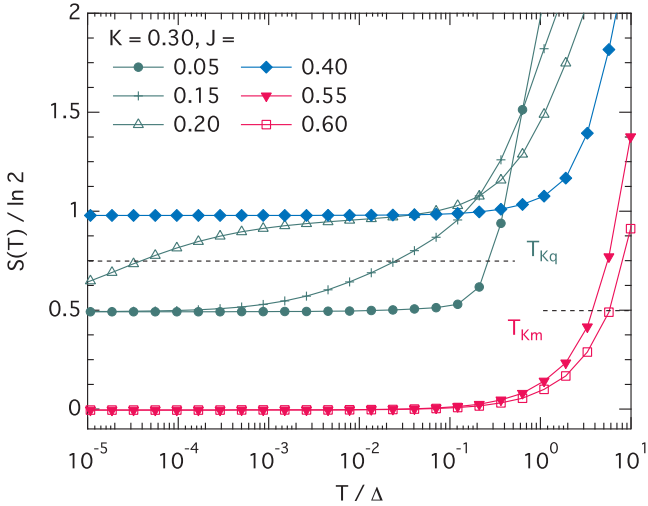}
}
\caption{(Color online) $T$ dependence of the impurity entropy for $K=0.30$. $T_{\rm Km}$ has a weak $J$ dependence in the FL phase, while $T_{\rm Kq}$ is strongly suppressed as $J$ increases in the NFL phase, yielding that the $\ln2$ entropy remains at the lowest $T$ in the FL+FS region.}
\label{S_K030}
\end{figure}

In Fig.~\ref{S_K030}, $S(T)$ is shown for $K=0.30$.
In this case, $T_{\rm Km}$ has a weak $J$ dependence in the FL phase, while $T_{\rm Kq}$ is strongly suppressed as $J$ increases in the NFL phase, yielding that the $\ln2$ entropy remains at the lowest $T$ in the FL+FS region.

\begin{figure}[t]
\centering{
\includegraphics[width=8.5cm]{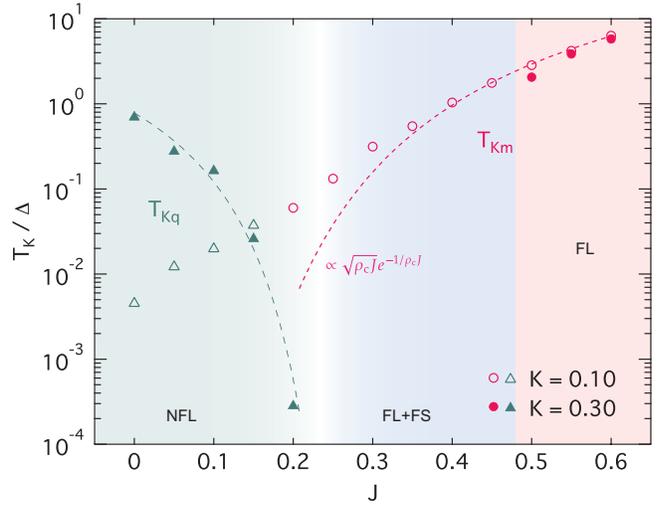}
}
\caption{(Color online) $T_{\rm Km}$ and $T_{\rm Kq}$, for $K=0.10$ and $K=0.30$. For larger $J$, $T_{\rm Km}$ follows the conventional formula of the Kondo temperature, $\sqrt{\rho_{c}J}\exp(-1/\rho_{c}J)$, where $\rho_{c}$ is the density of states of the conduction electrons at the Fermi energy. On the contrary, $T_{\rm Kq}$ is strongly suppressed for $K=0.30$ toward the FL+FS region.}
\label{TK}
\end{figure}

We summarize the $J$ dependences of $T_{\rm Km}$ and $T_{\rm Kq}$, for $K=0.10$ and $K=0.30$ in Fig.~\ref{TK}.
For larger $J$, $T_{\rm Km}$ follows the conventional formula of the Kondo temperature, $\sqrt{\rho_{c}J}\exp(-1/\rho_{c}J)$, where $\rho_{c}$ is the density of states of the conduction electrons at the Fermi energy.
$T_{\rm Kq}$ also increases monotonously as $J$ increases for $K=0.10$.
On the contrary, $T_{\rm Kq}$ is strongly suppressed due to the competition, and the residual $\ln2$ entropy remains in the FL+FS region, $0.2\lesssim J \lesssim 0.5$ for $K=0.30$.

\begin{figure}[t]
\centering{
\includegraphics[width=8.5cm]{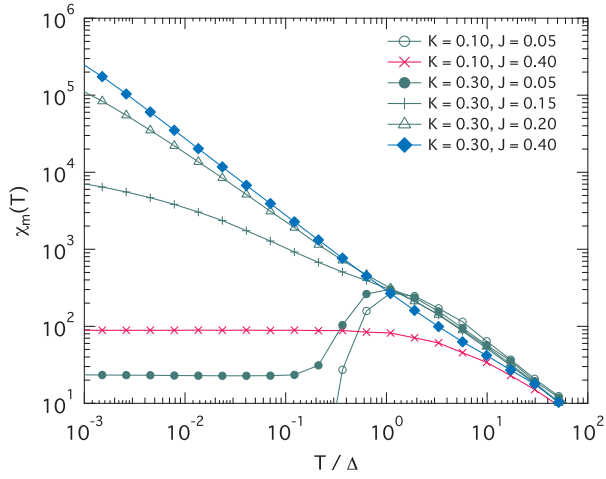}
}
\caption{(Color online) $T$ dependence of the magnetic susceptibility, $\chi_{\rm m}(T)$. It almost vanishes at temperatures below $\Delta$ for $K=0.10$, $J=0.05$, while it shows a typical FL behavior for $K=0.10$ and $J=0.40$. $\chi_{\rm m}(T)$ becomes a Curie-like behavior toward the FL-FS region for $K=0.30$ with increase of $J$.}
\label{chim}
\end{figure}

The magnetic property can be elucidated by the $T$ dependence of the impurity magnetic susceptibility, $\chi_{\rm m}(T)$.
We compute it as the expectation value of $S_{fz}$ under the Zeeman term, $-h_{s}S_{fz}$ with a small field $h_{s}=10^{-6}$.
In Fig.~\ref{chim}, $\chi_{\rm m}(T)$ almost vanishes at $T$ below $\Delta$ for $K=0.10$, $J=0.05$, while it shows a typical FL behavior below $T_{\rm Km}$ for $K=0.10$ and $J=0.40$.
On the other hand, $\chi_{\rm m}(T)$ shows the Curie-like behavior toward the FL-FS region for $K=0.30$ with increase of $J$.
This indicates the presence of the effective free magnetic spin, $S=1/2$.

\begin{figure}[t]
\centering{
\includegraphics[width=8.5cm]{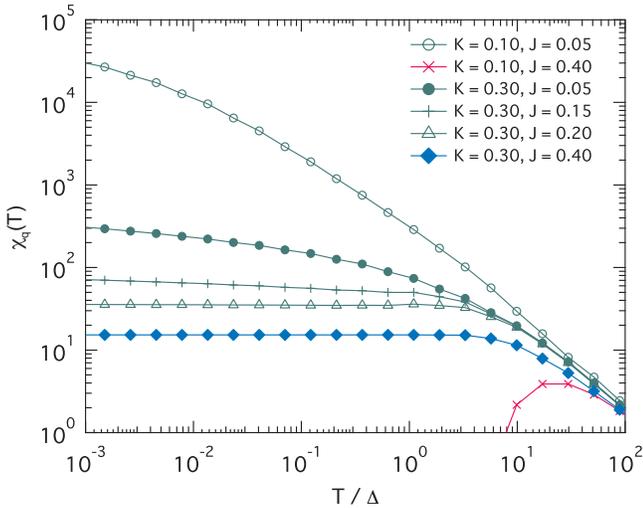}
}
\caption{(Color online) $T$ dependence of the quadrupolar susceptibility, $\chi_{\rm q}(T)$. For $K=0.10$ and $J=0.05$, $\chi_{\rm q}(T)$ shows a logarithmic divergence below $T_{\rm Kq}$, as a typical behavior of the two-channel Kondo model.
On the contrary, $\chi_{\rm q}(T)$ vanishes at $T$ below $T_{\rm Km}$ for $K=0.10$ and $J=0.40$.
For $K=0.30$, $\chi_{\rm q}(T)$ changes from the NFL behavior to FL behavior as $J$ increases toward the FL+FS region.}
\label{chiq}
\end{figure}

In the similar procedure, the $T$ dependence of the impurity quadrupolar susceptibility, $\chi_{\rm q}(T)$ is obtained by the expectation value of $T_{fz}$ in the presence of the $-h_{q}T_{fz}$ term with $h_{q}=10^{-6}$.
Figure~\ref{chiq} shows the $T$ dependence of $\chi_{\rm q}(T)$.
For $K=0.10$ and $J=0.05$, $\chi_{\rm q}(T)$ shows a logarithmic divergence below $T_{\rm Kq}$, as a typical behavior of the two-channel Kondo model.
On the contrary, $\chi_{\rm q}(T)$ vanishes at $T$ below $T_{\rm Km}$ for $K=0.10$ and $J=0.40$ where the magnetic Kondo effect dominates the low-energy phenomena.
For $K=0.30$, $\chi_{\rm q}(T)$ changes from the NFL behavior to FL behavior with the saturating $T$ dependence, as $J$ increases toward the FL+FS region.

\subsection{Excitation spectra}

Next, let us investigate the dynamical properties of the magnetic and quadrupolar excitations.
In the NRG, the spectral densities at $T=0$ are computed for the frequency $\omega=\alpha\,\Lambda^{-(N-1)/2}$ with $\alpha=2.0$ by means of the spectral representation of the two-particle Green's functions for $S_{fz}$ and $T_{fz}$.
In the expression of the spectral representation, we approximate the delta function as the logarithmic gaussian,
\begin{equation}
\delta(\omega-\epsilon)\to\frac{1}{\sqrt{\pi}\eta\epsilon}e^{-\eta^{2}/4}\exp\left[-\left(\frac{\ln(\omega/\epsilon)}{\eta}\right)^{2}\right],
\quad
(\epsilon>0),
\end{equation}
with the broadening factor, $\eta=0.7$.

\begin{figure}[t]
\centering{
\includegraphics[width=8.5cm]{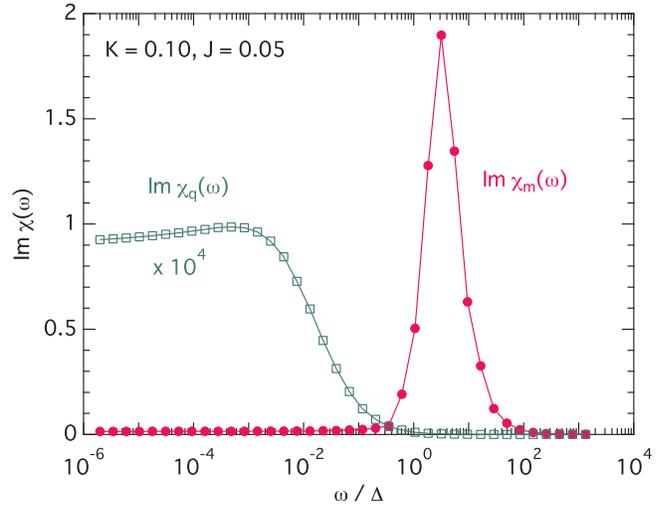}
}
\caption{(Color online) ${\rm Im}\,\chi_{\rm m}(\omega)$ and ${\rm Im}\,\chi_{\rm q}(\omega)$ for $K=0.10$ and $J=0.05$.}
\label{chi_K010J005}
\end{figure}

\begin{figure}[t]
\centering{
\includegraphics[width=8.5cm]{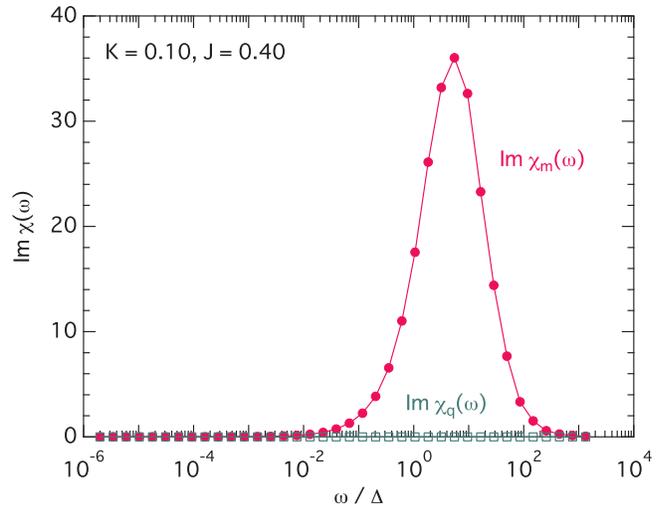}
}
\caption{(Color online) ${\rm Im}\,\chi_{\rm m}(\omega)$ and ${\rm Im}\,\chi_{\rm q}(\omega)$ for $K=0.10$ and $J=0.40$.}
\label{chi_K010J040}
\end{figure}

In the NFL phase, the excitation spectra are shown in Fig.~\ref{chi_K010J005} for $K=0.10$, $J=0.05$.
${\rm Im}\,\chi_{\rm m}(\omega)$ shows the typical peaked structure at $\omega\sim T_{\rm Km} \gtrsim\Delta$ corresponding to the magnetic Kondo effect.
On the other hand, ${\rm Im}\,\chi_{\rm q}(\omega)$ exhibits the NFL excitation spectrum (${\rm Im}\,\chi_{\rm q}(\omega)\propto\ln\omega$), which is consistent with the logarithmically divergent quadrupolar susceptibility, $\chi_{\rm q}(T)$ as shown in Fig.~\ref{chiq}.
In the FL phase for $K=0.10$, $J=0.40$, ${\rm Im}\,\chi_{\rm q}(\omega)$ is completely suppressed, and only the magnetic excitation with the peak at $\omega\sim T_{\rm Km}$ is prominent as shown in Fig.~\ref{chi_K010J040}.

\begin{figure}[t]
\centering{
\includegraphics[width=8.5cm]{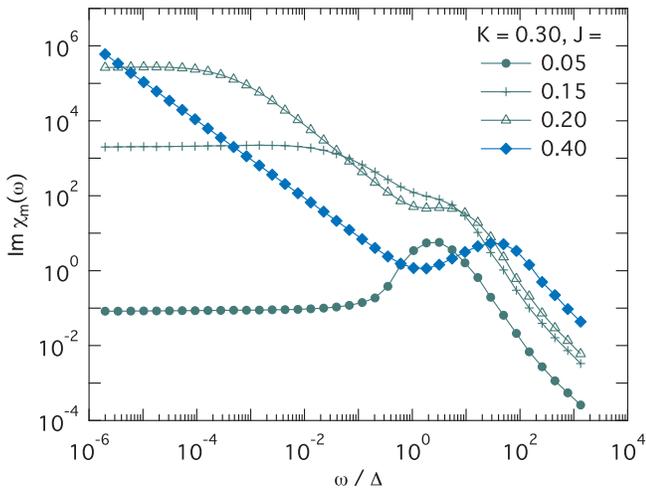}
}
\caption{(Color online) ${\rm Im}\,\chi_{\rm m}(\omega)$ for $K=0.30$. As $J$ increases, the ${\rm Im}\,\chi_{\rm m}(\omega)$ changes from the peaked structure to the divergent behavior ${\rm Im}\,\chi_{\rm m}(\omega)\propto 1/\omega$ with the higher energy-shift of the peak.}
\label{chim_K030}
\end{figure}

\begin{figure}[t]
\centering{
\includegraphics[width=8.5cm]{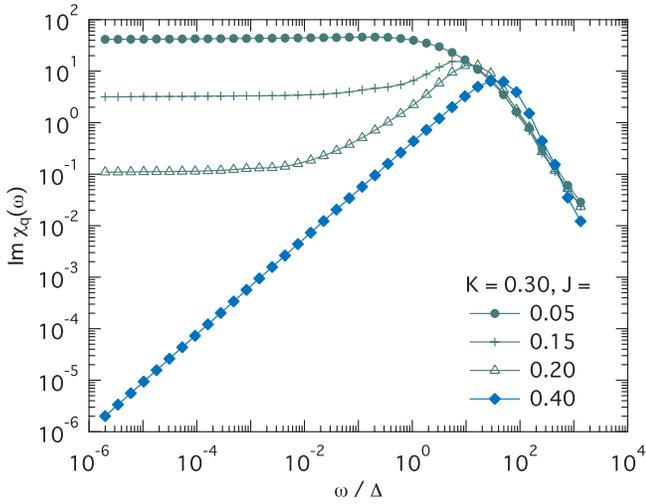}
}
\caption{(Color online) ${\rm Im}\,\chi_{\rm q}(\omega)$ for $K=0.30$. The ${\rm Im}\,\chi_{\rm q}(\omega)$ changes from the NFL behavior to the FL behavior as ${\rm Im}\,\chi_{\rm q}(\omega)\propto\omega$.}
\label{chiq_K030}
\end{figure}

In the case of the competing Kondo effects, Figs.~\ref{chim_K030} and \ref{chiq_K030} show the spectral intensities of the magnetic and quadrupolar susceptibilities for $K=0.30$.
As $J$ increases, ${\rm Im}\,\chi_{\rm m}(\omega)$ changes from the peaked structure as similar to that in Fig.\ref{chi_K010J005} to the divergent behavior ${\rm Im}\,\chi_{\rm m}(\omega)\propto 1/\omega$ with the higher energy-shift of the peak.
The divergent behavior in the latter reflects the Curie-like susceptibility.
On the contrary, ${\rm Im}\,\chi_{\rm q}(\omega)$ changes from the NFL behavior as in Fig.~\ref{chi_K010J005} to the FL behavior as ${\rm Im}\,\chi_{\rm q}(\omega)\propto\omega$.

Finally, we discuss the single-particle excitation spectra.
For the extended two-channel Kondo model, we introduce the dimensionless current operator for $\sigma=\uparrow,\downarrow$ as
\begin{align}
j_{\sigma}&\equiv\frac{1}{\sqrt{J^{2}+K^{2}}}\sum_{km}[c_{km\sigma}^{},H-H_{c}]
\cr&
=\frac{J}{2\sqrt{J^{2}+K^{2}}}\sum_{km}(\sigma S_{fm}^{z}c_{km\sigma}+S_{fm}^{-\sigma}c_{km-\sigma})
\cr&\quad\quad
+\frac{K}{2\sqrt{J^{2}+K^{2}}}\sum_{km}(mT_{f}^{z}c_{km\sigma}+T_{f}^{m}c_{km\sigma}),
\end{align}
where we adopt the signs in the expression as $\uparrow\to +$, $\downarrow\to-$, $m=1\to +$, and $m=2\to -$, respectively.
The $T$-matrix in the spin $\sigma$ channel is given by the retarded Green's function for the current operator as
\begin{equation}
T_{\sigma}(\omega)=i\int_{0}^{\infty}dt\,e^{i\omega t}\Braket{\left\{j_{\sigma}^{}(t),j_{\sigma}^{\dagger}\right\}}.
\end{equation}
The $T$-matrix is proportional to the self-energy of the conduction electrons, $\sum_{m}\Sigma_{m\sigma}(\omega)$, and its imaginary-part can be related to the ``$f$-electron density of states''.
It is also useful to deduce a development of the Kondo resonance peak and the temperature dependence of the resistivity from the imaginary part of the $T$ matrix, $T''(\omega)\equiv{\rm Im}\,T(\omega)$.

\begin{figure}[t]
\centering{
\includegraphics[width=8.5cm]{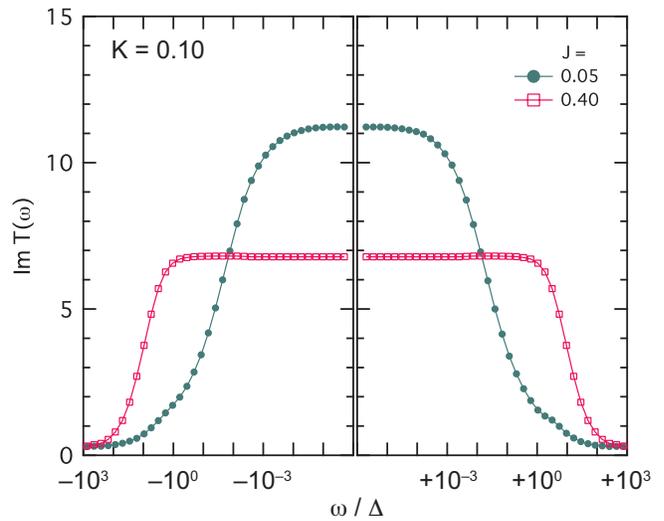}
}
\caption{(Color online) ${\rm Im}\,T(\omega)$ for $K=0.10$. For $J=0.05$ ($J=0.40$), it follows $\propto {\rm const.} -|\omega|^{1/2}$ (${\rm const.} -\omega^{2}$) that indicates that the resistivity from the $T=0$ value behaves as $\Delta\rho\propto T^{1/2}$ ($T^{2}$).}
\label{T_K010}
\end{figure}

\begin{figure}[t]
\centering{
\includegraphics[width=8.5cm]{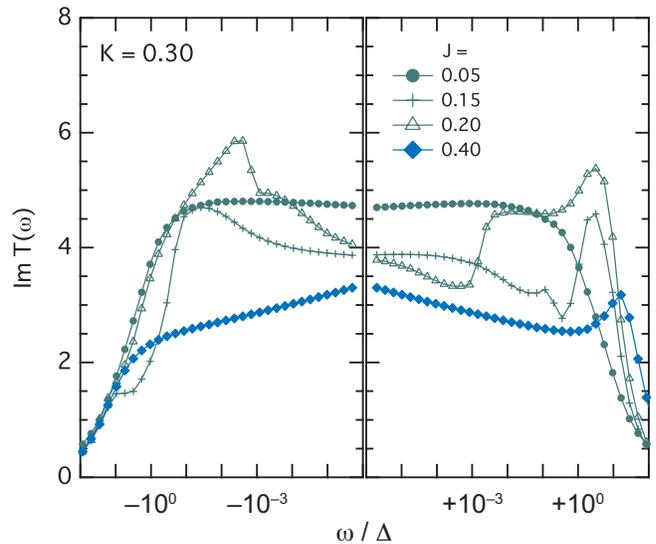}
}
\caption{(Color online) ${\rm Im}\,T(\omega)$ for $K=0.30$. As $J$ increases, it changes from the NFL behavior $\propto {\rm const.} -|\omega|^{1/2}$ to more singular behavior, $\propto {\rm const.} -\ln|\omega|$.}
\label{T_K030}
\end{figure}

Figures~\ref{T_K010} and \ref{T_K030} show the prime examples of the spectra in the NFL and FL phases, and the transition from the NFL to the FL+FS state, respectively.
In any cases, the Kondo resonance peak develops at low energies resulting from the magnetic and/or quadrupolar Kondo effects.
In Fig.~\ref{T_K010}, $T''(\omega)\propto {\rm const.}-|\omega|^{1/2}$ for $J=0.05$ in the NFL phase, while $T''(\omega)\propto {\rm const.}-\omega^{2}$ for $J=0.40$ in the FL phase.
It is interesting to note that the Kondo resonance peak appears even in the FL+FS region, where the magnetic spin looks as if it is free with the $\ln 2$ entropy.
In the case of the NFL and FL phases as shown in Fig.~\ref{T_K010}, the symmetric spectra are obtained due to the particle-hole symmetry.
For $K=0.30$ and small $J$, the spectrum is similar to that of the NFL.
However, it changes drastically to much more singular behavior, $T''(\omega)\propto {\rm const.}-\ln|\omega|$ as $J$ increases toward the FL+FS region.
The logarithmic divergence in $T''(\omega)$ is cut off eventually at the order of $T_{\rm Kq}$ which is remarkably small in the FL+FS region.
It can be deduced the temperature dependence of the resistivity $\rho(T)$ from the $T''(\omega)$, i.e., it may be obtained by replacing $\omega$ by $T$ in $T''(\omega)$.

\subsection{Relevance to Pr$T_{2}X_{20}$}

Now, let us discuss a relevance to the prominent observations in Pr$T_{2}X_{20}$ systems based on the results obtained in the previous sections, although our analysis only takes account of the single-impurity Kondo effects, and a lattice effect may alter coherent properties, and lead to long-range orders.

\begin{figure}[t]
\centering{
\includegraphics[width=8.5cm]{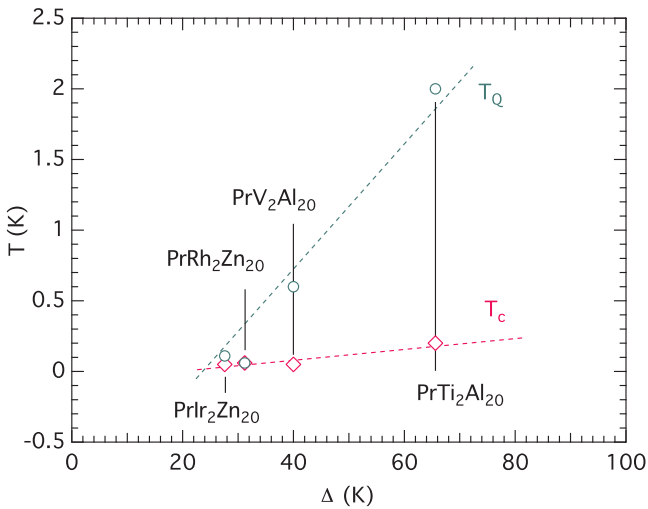}
}
\caption{(Color online) Quadrupole ordering temperature, $T_{\rm Q}$ and the superconducting $T_{\rm c}$ versus the 1st CEF excitation energy, $\Delta$ for Pr$T_{2}X_{20}$. The monotonic $\Delta$ dependence of $T_{\rm Q}$ and $T_{\rm c}$ indicates that the overall energy scale increases in the order of PrIr$_{2}$Zn$_{20}$ (PrRh$_{2}$Zn$_{20}$), PrV$_{2}$Al$_{20}$, and PrTi$_{2}$Al$_{20}$. It is expected that the hybridization strengths also increase in the same order of $\Delta$.}
\label{T_Delta}
\end{figure}

Figure~\ref{T_Delta} shows the $\Delta$ dependence of the quadrupole ordering temperature $T_{\rm Q}$, and the superconducting $T_{\rm c}$ for extensively studied four Pr 1-2-20 systems.~\cite{Onimaru10,Onimaru11,Sakai11,Ishii11,Sakai12,Sato12,Onimaru12,Iwasa13,Tsujimoto14,Araki14}
The monotonous increasing behaviors of $T_{\rm Q}$ and $T_{\rm c}$ with increase of $\Delta$ indicates that the overall energy scale increases in order of PrIr$_{2}$Zn$_{20}$ (PrRh$_{2}$Zn$_{20}$), PrV$_{2}$Al$_{20}$ and PrITi$_{2}$Al$_{20}$.
Naively, it is also expected that the hybridization strengths or the Kondo couplings increase in the same order of the overall energy scale.
Note that however since the quadrupolar and magnetic Kondo coupling comes from the same hybridization channels of the $\Gamma_{8}$ orbitals, the ratio $J/K$ would be the order of unity.

In PrIr$_{2}$Zn$_{20}$ and PrRh$_{2}$Zn$_{20}$, the resistivity decreases monotonously from the room temperature, and there is no logarithmic increase at higher temperatures above $\Delta$, meaning that no magnetic Kondo effect takes place in these compounds.
Below 1 K and above $T_{\rm Q}$, they exhibit the unconventional behaviors in the resistivity, specific heat, and Seebeck coefficient.~\cite{Onimaru10,Onimaru11,Onimaru12,Ikeura14,Machida15,Ishii13,Matsumoto15,Onimaru16}
The most remarkable feature of these compounds is that the $T$ dependence of the resistivity and specific heat under small magnetic fields satisfy a scaling relation.~\cite{Ikeura14,Machida15,Onimaru16a}
Namely, they can be described by a single curve as a function of $T/T_{\rm QKL}$ for several values of the magnetic fields where $T_{\rm QKL}$ is the characteristic temperature.
The universal curves coincide well with the theoretical predictions of the quadrupolar Kondo (lattice) model.~\cite{Cox98,Tsuruta15}
This fact strongly suggests that the lattice quadrupolar Kondo effect is indeed realized in PrIr$_{2}$Zn$_{20}$ and PrRh$_{2}$Zn$_{20}$.
The situation for these compounds seems to correspond to the small $J$ and $K$ region in our phase diagram, Fig.~\ref{phase}, and the quadrupolar Kondo effect predominates over the magnetic Kondo effect.

On the contrary, PrV$_{2}$Al$_{20}$ and PrTi$_{2}$Al$_{20}$ exhibit the logarithmic increases in the resistivity above 60 K, indicating the occurrence of the magnetic Kondo effect.
The former shows the NFL behaviors similar to those of PrIr$_{2}$Zn$_{20}$ below 3 K, while the latter exhibits the ordinary $T^{2}$ dependence below 20 K.~\cite{Sakai11,Sakai12,Tsujimoto14,Araki14,Koseki11,Matsubayashi12,Matsubayashi14}
Thus, it is important to test whether the universal scaling relation is also valid for PrV$_{2}$Al$_{20}$ or not in order to elucidate a universal feature of the quadrupolar Kondo effect.~\cite{Matsubayashi16}
For PrTi$_{2}$Al$_{20}$ on the other hand, applying pressure up to 8.7 GPa, the resistivity in PrTi$_{2}$Al$_{20}$ shows a drastic change from the FL behavior to NFL behavior above about 3 GPa, and this NFL behavior is quite different from that in PrIr$_{2}$Zn$_{20}$ and PrRh$_{2}$Zn$_{20}$.
The change of the resistivity behaviors is very sensitive to the pressure.~\cite{Matsubayashi12,Matsubayashi14,Matsubayashi16}
$T_{c}$ shows a peak-like structure at the lowest $T_{\rm Q}$ with a sizable mass enhancement.
Interestingly, at ambient pressure the entropy of PrV$_{2}$Al$_{20}$ is gradually decreasing from $R\ln2$ on cooling below 5 K, while the entropy of PrTi$_{2}$Al$_{20}$ stays at about $R\ln 2$ above 2 K and then it decreases abruptly below 2 K.~\cite{Tsujimoto14}
Although this behavior indicates the presence of robust degeneracy of the localized degrees of freedom, the development of the Kondo resonance peak at the Fermi energy has been observed by the Pr $3d$-$4f$ resonant photoemission spectroscopy.~\cite{Matsunami11}
All of the facts (i) the robust $\ln2$ entropy, (ii) the drastic crossover between the NFL and FL behaviors being quite sensitive to the pressure, and (iii) the presence of the magnetic Kondo effect suggest a relevance to the peculiar state in the FL+FS region.
This is because the Ti-system has a larger energy scale and then the possible competition of two Kondo effects may arise.
A detailed investigation along this scenario is left for future study based on a lattice version of the extended two-channel Kondo model.
For experimental side, investigations on the magnetic and dynamical properties are highly desired under high pressures.

\section{Summary}

We have discussed possible competition between the quadrupolar ($K$) and magnetic ($J$) Kondo effects on the basis of the extended two-channel Kondo model.
There are three distinct states in the model, (i) the (local) Fermi-liquid state ($J\gg K$), (ii) the (local) non-Fermi-liquid state ($J\ll K$), and (iii) the marginally free-spin state ($J\approx K$) with peculiar thermodynamic and dynamical properties.
The candidate compounds for this competition are the recently investigated Pr$T_{2}X_{20}$.
The low-temperature NFL behaviors in PrIr$_{2}$Zn$_{20}$ and PrRh$_{2}$Zn$_{20}$ seem to be the occurrence of the pure quadrupolar Kondo effect as the state (ii) with small $J$ and $K$.
The similar NFL behaviors observed in PrV$_{2}$Al$_{20}$ may be interpreted in a similar way as the Zn-systems in which the energy scale of PrV$_{2}$Al$_{20}$ is roughly twice larger than PrIr$_{2}$Zn$_{20}$.
The system seems to be located in the region of $K<0.15$ in Fig.~\ref{phase} where the state (iii) does not appear as $J$ increases.
On the contrary, PrTi$_{2}$Al$_{20}$ seems to be the case that the crossover between the NFL and FL+FS may occur by applying the pressure since the energy scale of PrTi$_{2}$Al$_{20}$ is considerably larger than the other systems.
To reveal intrinsic characteristics of PrTi$_{2}$Al$_{20}$,~\cite{Onimaru16} further investigations especially on the magnetic and dynamical properties are highly desired under high pressures.
Extensive investigations for corresponding dilute Pr systems~\cite{Onimaru11,Matsumoto15} could bring about further insights on the peculiar Kondo effects in the non-Kramers Kondo systems.

\section*{Acknowledgments}
The author would like to thank K. Izawa, T. Onimaru, K. Umeo, K. Matsubayashi, and S. Nakatsuji for fruitful discussions and for showing him the latest experimental data prior to publication.
This work was supported by JSPS KAKENHI Grant Number 15K05176, and 15H05885 (J-Physics).

\end{document}